\documentclass[preprint,showpacs,preprintnumbers,amsmath,amssymb]{revtex4}

\usepackage{graphicx}
\usepackage{dcolumn}
\usepackage{bm}
\usepackage{color}
\newcommand{\be}{\begin{equation}}
\newcommand{\ee}{\end{equation}}
\newcommand{\ba}{\begin{eqnarray}}
\newcommand{\ea}{\end{eqnarray}}

\begin{document}

\title{Self-dual formulations of $d=3$ gravity theories in the
path-integral framework}

\author{P.~J.~Arias$^{a}$}\email{pio.arias@ciens.ucv.ve}
\author{F.~A.~Schaposnik$^{b}$}\email{fidel@fisica.unlp.edu.ar}\thanks{Associated with CICBA}
\affiliation{$^a$Centro de F\'\i sica Te\'orica y Computacional\\
Facultad de Ciencias, Universidad Central de Venezuela\\
AP 20513, Caracas 1020-A, Venezuela.\\
$^b$Departamento de F\'\i sica, Universidad Nacional de La Plata/IFLP\\
C.C.67 - 1900 La Plata - Argentina.}

\begin{abstract}
We study the connection, at the quantum level, between $d=2+1$
dimensional self-dual models
with actions of growing  (from first to fourth) order, governing the dynamics of
helicity $\pm2$ massive excitations.   We obtain identities  between   generating functionals of
the different models  using the path-integral framework, this
allowing to establish dual maps among relevant vacuum  expectation values.
We check consistency of these  v.e.v.'s  with the
gauge invariance gained in each mapping.

\end{abstract}


\maketitle

\section{Introduction}
Since its introduction nearly three decades ago  \cite{DJT}
 $2+1$ dimensional topologically massive gravity, with a third order P and T odd   action  governing the dynamics of
a single massive graviton,  has attracted a lot of attention and
several dual formulations have been found \cite{ArKh}-\cite{Pio2}.
Indeed, various  $2+1$ dimensional models   describe, locally, a
single massive spin $2$ excitation, differing in their gauge
symmetries and sharing the common feature of  parity and time
reversal invariances violation. These properties come from the fact that in
$2+1$ dimensions the angular momentum tensor $\mathbb{M}^{\mu\nu}$
is dual to a pseudo-vector $\mathbb{J}^{\mu}$ so that the  Casimir operators of the Poincar\'e group
 are $\mathbb{P}_{\mu}\mathbb{P}^{\mu}=-m^2$
and $\mathbb{P}_{\mu}\mathbb{J}^{\mu}=ms$, with  $m$
  the mass and $s$ the spin of the excitation, {identified}
with its helicity.

Concerning P and T invariant
models, they must necessarily  have two excitations with the same
 mass
and opposite helicities\cite{binegar}-\cite{jackiwnair}. In fact,   a fourth order
unitary gravity model in which the graviton acquires mass without
the introduction of extra fields was recently  proposed \cite{BHT}-\cite{Nakasone}.
This
model is equivalent to the Fierz-Pauli model at the linear level
and in this sense it is P and T invariant and describes two
excitation of the same mass and opposite helicities.

The equations of motion of the single massive excitation models
relate the relevant field with its curl as it happens with
self-duality equations arising in vector
theories~\cite{PK}-\cite{DJself}. Indeed, the minimal description
of a spin 2  theory  leads to the equation of motion~\cite{ArKh}

\begin{equation}
\left(h^{Tt}\right)_\mu^a=-\frac{1}{m}{\varepsilon_{\mu}}^{\nu\lambda}\partial_{\nu}\left(h^{Tt}\right)_{{\lambda}}^a
\label{sefi}
\end{equation}
with $\left(h^{Tt}\right)_\mu^a$ a symmetric, transverse and traceless
tensor which
can be associated to the linear deviation of the dreibein
${e_{\mu}}^a$.
(We are considering a $2+1$  metric with signature $(-,+,+)$ and denote  $\mu,\nu$
and $a,b$ curved and flat indices respectively.).

Equation (\ref{sefi}) can be derived from a first order action
with no gauge symmetries,  in terms of the tensor field
${h_{\mu}}^a$. It is resemblant to
 the self-dual equation for a vector field $A_\mu$,
\begin{equation}
A_{\mu}=\frac{1}{m}{\varepsilon_{\mu}}^{\nu\lambda}\partial_{\nu}A_{\lambda}
\end{equation}
in  self-dual (Abelian) gauge theories.

One can extend the model with self-dual equation (\ref{sefi}),
which we  call SD1,   endowing the action with
 a  gauge symmetry in such a way that one ends with
  massive excitations without  breaking the aforementioned symmetry. The
first extension that one can envisage is the  {self-dual} {\emph {intermediate
model}} (SD2) introduced in \cite{ArKh} with a second order action having a
  symmetry that corresponds to the linearization of diffeomorphism
invariance. In growing order then comes the third order {\emph{linear topologically massive
model}}  \cite{DJT} (SD3), with a diffeomorphism invariant action
 also exhibiting   Lorentz invariance when
written in terms of the dreibeins. Finally there is the so-called
{\emph{new topological massive model}} (SD4) \cite{andringa},
\cite{dalmazinew} with a fourth order action which, in addition,
has a linear conformal invariance.

These four models that describe a single massive excitation of
helicity $+2$ or $-2$   can be connected via duality
transformations that incorporate the corresponding gauge
symmetries in passing from SD1 to
SD3~\cite{deserself},\cite{dalmazi},\cite{Pio2} and to
SD4~\cite{ask}. A link
 joining the SD4 model was also presented in \cite{dalmazinew} and in the opposite direction it can
be seen that fixing the gauge one can go from SD3 to SD1 models
\cite{deserself},\cite{jorgepio},\cite{PJA}. Such connections have
been in general established at the classical level. A quantum
analysis has been outlined in \cite{dalmazi},\cite{Pio2} for
connections SD1 to SD3 and to SD4 in \cite{dalmazinewb}, basically
analyzing propagators in connection with unitarity.

It  is the purpose of the
 present work to study the dual equivalence  between the   four
massive models SDI (with $\rm I=1, \ldots 4$) at the quantum level.
The  idea is to follow the approach developed in the study of  self-dual and
the topologically massive vector models, whose equivalence was established at the level of equations of motions in
refs.\cite{PK}-\cite{DJself} and discussed at the quantum level
within the path-integral approach in
\cite{Fradkin:1994tt}-\cite{LeGuillou}.

Basically, we shall
introduce successive ``interpolating actions"  depending on two  fields: the one
of the departure model and the
 dual model one. Then, we shall define the  generating functional associated to the interpolation actions in the form
 of a path-integral over the two fields.
 Integrating over one or the other fields the connection   between the generating functionals of the SDI  models
 can be established and, from it, the connections between quantum correlation functions follow.

The plan of the paper is the following: in
the next section we introduce the
actions governing the dynamics of the different models, analyze their  symmetries and exhibit their equivalence at the classical level,
showing that the
equations of motions for the SD1 to $SD4$ models can be derived from a
 single equation for a symmetric, transverse and
traceless tensor that can be identified in each case with the
linear deviation of the dreibein, the linear spin connection, the
linear Schouten tensor and the linear Cotton tensor respectively.

We introduce in section 3 the interpolating action for models SD1
and SD2  and define the associated generating functional in the
path-integral framework. Integrating on one or the other field
will allow to connect vacuum expectation values for SD1 and SD2
models. An analogous procedure is described in section 4, now for
connecting models SD1 and SD2 with  SD3. In section 5 we extend
the calculation in order to connect the previous models with the
new topological massive model SD4. Finally, in section 6 we
present our conclusions.

\section{Dualities at the classical level}
 We take
the fields ${h_{\mu}}^a$ as the linear deviation of  dreibeins $e_{\mu}^a$
\begin{equation}
{e_{\mu}}^a= {\delta_{\mu}}^a+\kappa{h_{\mu}}^a
\label{deviation}
\end{equation}
with $\kappa$  a parameter related to Newton constant $G_N$,
$\kappa^2 = 8\pi G_N$. To first order in $\kappa$ flat indices
$a,b$ in field $h$ can be replaced by $\mu,\nu$ curved ones; we shall
nevertheless maintain the distinction in order to keep track of
this two type of indices. The symmetric part of $h$ is connected
with the linear deviation of the metric
\begin{equation}
g_{\mu\nu}=\eta_{\mu\nu}+ \kappa H_{\mu\nu} \qquad \qquad {\rm with}
\qquad \qquad H_{\mu\nu}=h_{\mu\nu}+h_{\nu\mu}
\end{equation}

The dual of the torsion free spin connection is
\begin{equation}
e{\omega_{\mu}}^{a}=(e_{\mu b}{e_{\rho}}^{a} -
\frac{1}{2}{e_{\mu}}^{a}e_{\rho
b})\varepsilon^{\rho\nu\lambda}\partial_{\nu}{e_{\lambda}}^{b}
\end{equation}
In the linear approximation, the resulting linearized spin
connection, becomes
\begin{eqnarray}\label{seis}
{{\omega^{L}}_{\mu}}^{a}(h)& = &(\eta_{\mu b}\delta_{\rho}^{a} -
\frac{1}{2}\delta_{\mu}^{a}\eta_{\rho b})\varepsilon^{\rho\nu\lambda}\partial_{\nu}{h_{\lambda}}^b \nonumber\\
& = &\frac{1}{2}\delta_\lambda^a\varepsilon^{\lambda\nu\rho}
\left(\partial_{\nu}(h_{\mu\rho}+h_{\rho\mu})-\partial_{\mu}h_{\nu\rho})\right)\nonumber \\
 & \equiv &   {[{W_{\mu}}^{a}]_{b}}^{\lambda}
{h_{\lambda}}^{b} = {W_{\mu}}^{a}(h).
 \label{tres}
\end{eqnarray}
From this expression it is easy to verify that
\begin{equation}
\varepsilon_{abc}\varepsilon^{\mu\nu\lambda}{\delta_{\lambda}}^{c}{W_{\nu}}^b(h)=\varepsilon^{\mu\nu\lambda}\partial_{\nu}h_{\lambda
a} \label{con}
\end{equation}
The linearized Einstein tensor $G^{L\mu\nu}(h)$ is then given by
\begin{eqnarray}
G^{L\mu\nu}(h) & = & -\varepsilon^{\nu\rho\sigma}\partial_{\rho}{W_{\sigma}}^{c}(h)\delta^{\mu}_{c}
 =   -\varepsilon^{\mu\alpha\rho}\varepsilon^{\nu\beta\sigma}\partial_{\alpha}\partial_{\beta}h_{\rho\sigma}
=-\frac{1}{2}\varepsilon^{\mu\alpha\rho}\varepsilon^{\nu\beta\sigma}\partial_{\alpha}\partial_{\beta}(h_{\rho\sigma}
+h_{\sigma\rho}),
\end{eqnarray}
while the Cotton tensor can be written in the linear approximation as
\begin{eqnarray}
C^{L\mu\nu}&=&\varepsilon^{\mu\rho\lambda}\partial_{\rho}{S^{L\nu}}_{\lambda}
=-\frac{1}{2}\varepsilon^{\mu\alpha\rho}\varepsilon^{\mu\beta\sigma}
\partial_{\alpha}\partial_{\beta}(W_{\rho\sigma}(h)
+W_{\sigma\rho}(h))\label{cotton}
\end{eqnarray}
where ${S^{L\mu}}_{\nu}$ is the  linearized Schouten tensor
\begin{eqnarray}\label{diez}
{S^{L\mu}}_{\nu}&=&{{G}^{L\mu}}_{\nu}-\frac{1}{2}\delta^{\mu}_{\nu}G^L
=-(\delta^{\mu}_{c}\eta_{\nu\rho}-
\frac{1}{2}\delta^{\mu}_{\nu}\eta_{\rho
c})\varepsilon^{\rho\sigma\lambda}\partial_{\sigma}{{W}_{\lambda}}^c(h)\nonumber
\\
&=&-{[{W^{\mu}}_{\nu}]_c}^{\lambda}{{W}_{\lambda}}^c(h).
\end{eqnarray}

Under diffeomorphism transformations ${h_{\mu}}^a$ and
${\omega_{\mu}}^a$ change as
\begin{equation}
\delta_{\zeta}{h_{\mu}}^{a}=\partial_{\mu}\zeta^{a} \ \ \ , \ \ \
\delta_{\zeta}\omega_{\mu}^{La}=0
\end{equation}
Concerning  Lorentz transformations, they take the form
\begin{equation}
\delta_{l}{h_{\mu}}^{a}={\varepsilon^{a}}_{bc}l^{b}\delta_{\mu}^{c}
\ \ \ , \ \ \  \delta_{l}\omega_{\mu}^{a}=\partial_{\mu}l^{a}
\end{equation}
 and for conformal transformations one has
\begin{equation}
\delta_{\rho}{h_{\mu}}^{a}=\frac{1}{2}\rho{{\delta}_{\mu}}^{a} \ \
\ , \ \ \ {\delta}_{\rho}\omega_{\mu}^{La}=
\frac{1}{2}{\varepsilon_{\mu}}^{\nu\sigma}\partial_{\sigma}\rho\delta_{\nu}^{a}.
\end{equation}

In order to see the effect of these transformations on the different
objects that we are considering we can make a decomposition of the
field $h$ in its irreducible components
\begin{eqnarray}\label{descomp}
h_{\mu\nu} & = & (H^{Tt}_{\mu\nu} + \frac{1}{2}P_{\mu\nu}H^{T} +
\rho_{\mu}\rho_{\nu}H^{L} +
\rho_{\mu}h_{\nu}^{T} +\rho_{\nu}h_{\nu}^{T} )
+(\varepsilon_{\mu\nu\lambda}V^{T\lambda}
+ \varepsilon_{\mu\nu\lambda}\rho^{\lambda}V^{L}) \\
& \equiv & h^{S}_{\mu\nu} + h^{A}_{\mu\nu},\nonumber
\end{eqnarray}
with
\begin{equation}
P^{\mu}_{\nu} = \delta^{\mu}_{\nu} - \rho^{\mu}\rho_{\nu} \qquad ,
\qquad \rho_{\mu} = \frac{\partial_{\mu}}{\Box^{1/2}}.
\end{equation}
and
\begin{eqnarray}
h^{S}_{\mu\nu} = h^{S}_{\nu\mu} \qquad &,& \qquad h^{A}_{\mu\nu} = -h^{A}_{\nu\mu}, \nonumber  \\
H_{\mu}^{Tt\mu} = 0 \qquad &,& \qquad \partial_{\mu}H^{Tt\mu}_{\nu}=0,\nonumber \\
\partial_{\mu}h^{T\mu} = 0 \qquad &,& \qquad \partial_{\mu}V^{T\mu} = 0.
\end{eqnarray}
In decomposition  (\ref{descomp}) {$H^{Tt}_{\mu\nu}$} are the spin
2 components of the field $h_{\mu\nu}$. The spin 1 components are
${h_{\mu}}^{T}$ and $V_{\mu}^{T}$ and the spin 0 components are
$H^T, H^L$ and $V^L$. {$H^{Tt}_{\mu\nu}$} is  symmetrical,
transverse and traceless
 and it is invariant under  diffeomorphism, Lorentz
and conformal transformations. Under diffeomorphism the sensitive
components are $H^{L}, h_{\mu}^{T}$ and $V_{\mu}^{T}$, while under
Lorentz tranformations  the sensitive components are $V_{\mu}^{T}$
and $V^L$. Under conformal transformation only $H^{T}$ and $H^{L}$
change.

It is important to recall that the absence of the antisymmetrical
part of ${h_{\mu}}^{a}$ in a given action  is a sign of explicit Lorentz
invariance. Similarly, if the antisymmetrical part of its
${\omega_{\mu}}^{a}$ is also absent, the action has an
explicit conformal invariance.

 In terms of
irreducible components, the {linearized} Einstein tensor takes the form
\begin{equation}\label{eq:ap250}
G^{L}_{\mu\nu}= -\Box(H_{\mu\nu}^{Tt} -
\frac{1}{2}P_{\mu\nu}H^{T}),
\end{equation}
showing its explicit invariance under diffeomorphism and Lorentz
transformations. Concerning the Cotton tensor, it can be written as
\begin{equation}\label{eq:ap253}
{C^{L\mu\nu}}=-\Box\varepsilon^{\mu\alpha\beta}\partial_{\alpha}H_{\beta}^{Tt\nu}.
\end{equation}
making apparent its invariance under Lorentz and conformal
transformations, This tensor is also symmetric, transverse and
traceless.

Finally it is straightforward to see that when ${h_{\mu}}^a$ is
symmetric, transverse and traceless
\begin{equation}
h_{\mu\nu}=h_{\nu\mu} \ \ \ , \ \ \
{h_{\mu}}^a{{\delta}^{\mu}}_a=0 \ \ \ , \ \ \
\partial^{\mu}{h_{\mu}}^a=0
\end{equation}
 so are the linear spin connection, the linear Einstein tensor
and, trivially, the linear Cotton tensor.

Let us introduce a (2+1)-dimensional spin 2 excitation
$V^{Tt}_{\mu\nu}$ which afterwards will be  identified   either
with the spin 2 component of $h_{\mu\nu}$,  with the spin
connection $W_{\mu\nu}(h^{Tt})$, with the linearized Schouten
tensor $S_{\mu\nu}(h^{Tt})$ or with the linearized Cotton tensor
$C_{\mu\nu}(h)$.
 $V^{Tt}_{\mu\nu}$ is a symmetric, transverse and
traceless tensor satisfying the
equation \cite{binegar}-\cite{jackiwnair}
\begin{equation}\label{eqsd}
\frac{1}{2}{[(\mathbb{P}_{\mu}\mathbb{J}^{\mu}_{2}) -
ms\mathbb{I}]_{\alpha\beta}}^{\mu\nu}V^{Tt}_{\mu\nu}=0,
\end{equation}
with $s=\pm2$.
 {${\mathbb{P}}_{\mu}$} is the
momentum operator and  $\mathbb{J}^{\mu}_{2}$ is given
by~\cite{Pio2}
\begin{eqnarray}
{(\mathbb{J}_{2}^{\mu})_{\alpha\beta}}^{\gamma\sigma} & = &
-i{{\varepsilon^{\mu}}_{\lambda}}^{\rho}x^{\lambda}
{(\mathbb{I}_{s})_{\alpha\beta}}^{\gamma\sigma}\partial_{\mu}
+\frac{i}{2}(\delta^{\gamma}_{\alpha}{\varepsilon_{\beta}}^{\mu\sigma}
+ \delta^{\gamma}_{\beta}{\varepsilon_{\alpha}}^{\mu\sigma} +
+\delta^{\sigma}_{\alpha}{\varepsilon_{\beta}}^{\mu\gamma} +
\delta^{\sigma}_{\beta}{\varepsilon_{\alpha}}^{\mu\gamma}) \\
& \equiv & -{{\varepsilon^{\mu}}_{\lambda}}^{\rho}x^{\lambda}
{(\mathbb{P}_{\mu})_{\alpha\beta}}^{\gamma\sigma} +
{(j^{\mu})_{\alpha\beta}}^{\gamma\sigma},\label{j2}
\end{eqnarray}
with ${{(\mathbb{I}_{s})}_{\alpha\beta}}^{\gamma\sigma}=
\frac{1}{2}(\delta_{\alpha}^{\gamma}\delta_{\beta}^{\sigma} +
\delta_{\alpha}^{\sigma}\delta^{\gamma}_{\beta})$.
 One recognizes
the first term in (\ref{j2}) as the orbital part of the angular momentum.
These operators satisfy the Poincar\'e algebra
\begin{eqnarray}
i{[\mathbb{P}^{\mu},\mathbb{P}^{\nu}]_{\alpha\beta}}^{\rho\sigma}
& = & 0, \nonumber \\
i{[\mathbb{P}^{\mu},\mathbb{J}^{\nu}_{2}]_{\alpha\beta}}^{\rho\sigma}
& = &
-\varepsilon^{\mu\nu\lambda}{(\mathbb{P}_{\lambda})_{\alpha\beta}}^{\rho\sigma},\\
i{[\mathbb{J}^{\mu}_{2},\mathbb{J}^{\nu}_{2}]_{\alpha\beta}}^{\rho\sigma}
& = &
-\varepsilon^{\mu\nu\lambda}{(\mathbb{J}_{2\lambda})_{\alpha\beta}}^{\rho\sigma}.\nonumber
\end{eqnarray}

Equation (\ref{eqsd}) can be written in the form
\begin{equation}
-{\varepsilon_{\mu}}^{\lambda\rho}\partial_{\lambda}V_{\rho\nu}^{Tt}
\mp mV_{\mu\nu}^{Tt}=0 \label{eq. s2}
\end{equation}
or, using (\ref{tres}),
\begin{equation}
V_{\mu\nu}^{Tt}=\mp\frac{1}{m} W_{\mu\nu}^{Tt}(V) \;,
\label{dro}
\end{equation}
which we recognize as the self dual equation for
$V_{\mu\nu}^{Tt}$~\cite{ArKh}-\cite{deserself}. This equation is
sensitive to $P$ and $T$ transformations. $V^{Tt}$ has two
independent degrees of freedom but one can see that only one
combination propagates~\cite{ArKh},\cite{PJA}.

 If $V^{Tt}_{\rho\sigma}$  is
identified with  the spin 2 component of the linearization of the
dreibein $h_{\mu\nu}$, then equation (\ref{dro}) coincides, after
eliminating the spurious degrees, with the equation of motion of
the self-dual (SD) model with action~\cite{ArKh}
\begin{eqnarray}
S_{SD1} & = &\mp\frac{m}{2}\int d^{3}x\,[h_{\mu\rho}
\varepsilon^{\mu\nu\lambda}\partial_{\nu}{h_{\lambda}}^{\rho}
\pm{m}(h_{\mu\nu}h^{\nu\mu} -h_{\mu}^{\mu}h_{\nu}^{\nu})] \\
& \equiv & \frac{1}{2}\int d^{3}x\, {h_{\mu}}^{a}{{{{K^{\pm}}_a}^
{\mu}}_{b}}^{\lambda} {h_{\lambda}}^{b}.  \label{sd1action}
\end{eqnarray}
The $\pm$ signs are related to the spin $\pm 2$ of the
excitacions~\cite{piorolando}; note that changing $m$ to $-m$
implies passing from the $+2$ to the $-2$ description. Action
(\ref{sd1action}) does not have any gauge invariance and the
antisymmetric part of the field ${h_{\mu}}^{a}$ acts as a
auxiliary field ensuring the spin 2 content. Taking $h$ to be
symmetric from the start one can see that there will be a spin 1
ghost remaining, associated with $\partial^{\mu}{H_{\mu\nu}}$
which propagates with mass $2m$ \cite{Pio2}.

There are other symmetric, transverse and traceless candidates
satisfying equation(\ref{eqsd}) They are the spin connection
${W_{\mu}}^a(H^{Tt})$, the Schouten tensor
${S_{\mu}}^{\nu}(H^{Tt})$ and the Cotton tensor
${C_{\mu}}^{\nu}(H^{Tt})$ written in terms of a symmetric,
transverese and traceless dreibein.
 When $V_{\mu\nu}^{Tt}$ is identified with the connection,  equation (\ref{eqsd}) coincides
 with the equations of motion, on the physical modes, of the so called  intermediate,
 (second order) action $S_{SD2}$ \cite{ArKh}
\begin{eqnarray}
S_{SD2}^{\pm}&=&\frac{1}{2}\int
d^{3}x \left({h_{\mu}}^{a}\varepsilon^{\mu\nu\lambda}
\partial_{\nu}W_{\lambda a}(h)\pm
m{h_{\mu}}^{a}\varepsilon^{\mu\nu\lambda}\partial_{\nu}h_{\lambda
a}\right)  \label{sd2action} \\
&=& \frac{1}{2}\int d^{3}x \left[{h_{\mu}}^{a}{{{{K^{\pm}}_a}^
{\mu}}_{b}}^{\lambda}\left( \mp\frac{1}{m}{W_{\lambda}}^b(h)\right)\right]\nonumber\\
&\equiv& S_E \pm S_{TCS} \; , \\
\label{sd2action2}
\end{eqnarray}
where ${{{{K^{\pm}}_a}^ {\mu}}_{b}}^{\lambda}$ is the same
evolution operator that appears in (\ref{sd1action}), and we have
identified the terms $S_E$ and $S_{TCS}$ with the Einstein action
term and the triadic Chern-Simons term \cite{AAKh} respectively.
The Einstein term can be written in terms of the symmetric part of
${h_{\mu}}^{a}$ so the contribution of the antisymmetric part in
the second term is just to ensure the non propagation of the spin
1 component of $H_{\mu\nu}$. This action corresponds to the
linearization of the so called Massive vector Chern-Simons
gravity~\cite{VCS}.

The intermediate action $S_{SD2}$ is invariant under
diffeomorphism transformations up to a surface term and it can be
proved, at the classical level, that it is related with the SD1
action $S_{SD1}$ after fixing the
gauge\cite{deserself},\cite{jorgepio},\cite{PJA}. It should be
noticed, however, that the space of classical solutions of the two
models is different (solutions have different topological
properties and those associated with action  $SD2$ include the non
trivial solution $\omega_{\mu\nu}(h)=0$ \cite{PJA}\cite{anyong},
absent in the SD1 model).

If one identifies   $V_{\mu\nu}^{Tt}$ with the linearized Schouten
tensor, eq.(\ref{eqsd})  becomes the equations of motion for
physical modes of the linear topologically massive action
\cite{DJT} ($S_{SD3}$)
\begin{eqnarray}
S^{\pm}_{SD3} & = & -\frac{1}{2}\int d^{3}x
\left[{h_{\mu}}^{a}\varepsilon^{\mu\nu\lambda}
\partial_{\nu}W_{\lambda a}(h)\pm
\frac{1}{2m}{W_{\mu}}^{a}(h)\varepsilon^{\mu\nu\lambda}\partial_{\nu}{W_{\lambda}}^{b}\right]\label{sd3action}
\nonumber\\
 &=&-S_E \pm S_{CS} \nonumber\\
&=& \frac{1}{2}\int d^{3}x \left[
\left(\mp\frac{1}{m}{W_{\mu}}^{a}(h)\right){{{{K^{\pm}}_a}^
{\mu}}_{b}}^{\lambda} \left(\mp\frac{1}{m}{W_{\lambda}}^b(h)\right)\right]\nonumber\\
&=& \frac{1}{2}\int d^{3}x\,\left[{h_{\mu}}^{a}{{{{K^{\pm}}_a}^
{\mu}}_{b}}^{\lambda}\left(-\frac{1}{m^2}{S_{\lambda}}^b(h)\right)\right]\label{sd3action2}
\end{eqnarray}
We have identified the Einstein term with a minus sign in front so
that it leads  to the correct propagation of its physical modes
\cite{DJT}. We have also identified the linear gravitational
Chern-Simons term, $S_{CS}$. The expression in the last line,
written in terms of the Schouten tensor will be useful later when
we make the connection with the other models at the quantum level.

Action $S^{\pm}_{SD3}$ is explicitly Lorentz invariant because it
depends only on the symmetrical part of $h_{\mu\nu}$ and it also
exhibits diffeomorphism invariance up to a surface term. It
can be written in terms of the symmetric part of $h$
\begin{eqnarray}
S_{SD3}&=& \frac{1}{4}\int d^3 x
\left[H_{\mu\nu}{G^L}^{\mu\nu}(H)\pm\frac{1}{m}H_{\mu\nu}{C^L}^{\mu\nu}(H)\right]
\label{sd3actionsim}\\
 &=&\frac{1}{4}\int d^3 x
\left[H_{\mu\alpha}{{K^{\pm}}^
{\mu\alpha}}_{\rho\beta}(-\frac{1}{m^2}{S^{\rho\beta}}(H))\right].
\label{sd3actionsim2}
\end{eqnarray}

It has been shown that SD3 connects with the SD2  model after a
gauge fixing~\cite{deserself},\cite{jorgepio},\cite{PJA}. It
should be noted that   the classical space of solutions of SD2 and
SD3 models is  different  because the later's one includes the
nontrivial solutions of the equation
$G^{\mu\nu}(h)=0$~\cite{deseranyon},
\cite{jorgepio}-\cite{anyong}.

Finally if we identify ${V^{Tt}}_{\mu\nu}$ in (\ref{eqsd}) with
the linear Cotton tensor, which has the same symmetries, the
equation correspond to the linearization of the so called new
topological massive model~\cite{andringa},\cite{dalmazinew}
\begin{eqnarray}
S_{SD4}&=&\frac{1}{4}\int d^3 x
\left[-\frac{1}{m^2}H_{\mu\alpha}\varepsilon^{\mu\nu\lambda}\partial_{\nu}{{C^{L}}_{\lambda}}^{\alpha}
\mp\frac{1}{m}H_{\mu\alpha}{C^L}^{\mu\alpha}\right]\label{sd4action}\\
&=&\frac{1}{4}\int d^3 x \left[H_{\mu\alpha}{K^{\pm}}^
{\mu\alpha,\rho\beta}\left(\frac{1}{m^3}{C_{\rho\beta}}(H)\right)\right].
\label{sd4action2}
\end{eqnarray}
This action is invariant under diffeomorphism and conformal
transformations. The sign in front of the first term in
(\ref{sd4action}) is so in order to have the correct description
of a massive excitation with helicity $\pm2$ depending on the sign
of the second term~\cite{andringa},~\cite{dalmazinewb}. This model
can be connected by duality transformation with the other SD
models~\cite{dalmazinew}.

We have then reviewed the connection, at the classical level, of
models with actions $S_{SD1}, S_{SD2}$, $S_{SD3}$ and $S_{SD4}$
defined in eqs.(\ref{sd1action}), (\ref{sd2action2}),
(\ref{sd3action2}) and (\ref{sd4action2}) respectively. The
connections between these models have been established in various
ways and in this sense they are considered as dual
models~\cite{ArKh},~\cite{deserself},~\cite{dalmazi},~\cite{Pio2},~\cite{dalmazinew},~\cite{dalmazinewb}.
We can   summarize these classical dualities as follows
\begin{equation}
S_{SD1}\;\;\;\;\Leftrightarrow \;\;\;\;S_{SD2} \;\;\;\;
\Leftrightarrow \;\;\;\; S_{SD3} \;\;\;\; \Leftrightarrow
\;\;\;\; S_{SD4}\label{graph} \end{equation}
As stressed in the introduction, the goal of our paper is to
analyze dualities at the quantum level using the path-integral
approach. Indeed, we shall connect in the following sections the
four models at the generating functional level, this allowing to
establish identities among correlation functions of the four SDI
model's partition functions. That is, connection (\ref{graph})
between classical actions will become one between generating
functionals. For this purpose we will introduce different {\emph
{interpolating actions}} that allows to pass from one model to the
other.

\section{Dualities at the quantum level:  the path-integral \\approach}
Following the ideas in \cite{PK}-\cite{LeGuillou}, let us
introduce the \emph{interpolating action} $S_I[ h,{\cal H}]$
\begin{equation}
S_I[ h,{\cal H}]=\frac{1}{2}\int d^{3}x(m{{\cal
H}_{\mu}}^{a}\varepsilon^{\mu\nu\lambda}\partial_{\nu}{\cal
H}_{\lambda
a}-2m{h_{\mu}}^{a}\varepsilon^{\mu\nu\lambda}\partial_{\nu}{\cal
H}_{\lambda a}
-{m^2}\varepsilon_{abc}\varepsilon^{\mu\nu\lambda}{h_{\mu}}^{a}{h_{\nu}}^{b}{\delta_{\lambda}}^{c})\label{SD0}
\end{equation}
where $h_{\mu}^{a}$ is the dreibein deviation field as defined in
(\ref{deviation}) and  ${\cal H}_{\mu}^{a}$   the corresponding
dual field.

Action $S_I$ is invariant, up to a surface term, under the
gauge transformations
\begin{equation}
\delta {{\cal H}_{\mu}}^{a}=\partial_{\mu}\xi^a \quad,\quad \delta
{h_{\mu}}^{a}=0,
\end{equation}
which reminds us of the linearized diffeomorphism transformation
in ${{\cal H}_{\mu}}^{a}$. The signs in the different terms of the
action corresponds in our conventions to a $+2$ helicity
description. The description for a $-2$ helicity is obtained
changing $m \to -m$. We will consider the positive helicity case
from here on.

Within the path integral framework  the partition function  $Z_I$
associated to action $S_I[ h,{\cal H}]$ reads
\begin{equation}
Z_I=\int{\cal D}{{\cal H}_{\mu}}^{a}{\cal
D}{h_{\mu}}^{a}\exp(iS_I[h,{\cal H}]). \label{ZI}
\end{equation}
We shall define the associated generating function $Z_I[j]$ by
coupling the $h$-field minimally to an external source $j_a^{\,\mu}$
\begin{equation}
Z_I[j]=\int{\cal D}{{\cal H}_{\mu}}^{a}{\cal
D}{h_{\mu}}^{a}\exp(iS_I[h,{\cal H};j]) \label{ZIJ}
\end{equation}
with
\begin{equation}
S_I[h,{\cal H};j] = S_I[h,{\cal H}] + \int d^3x {h_\mu}^a j_a^{\,\mu}
\end{equation}

Now, generating  functional $Z_I[j]$ as defined in (\ref{ZIJ}) can be connected with the generating
 functional for correlation functions
for   $h$-field with dynamics governed by the self-dual (SD1)
action. To see this, let us perform the path integral on ${{\cal
H}_{\mu}}^{a}$ in (\ref{ZIJ}). To this end, consider the ${{\cal
H}_{\mu}}^{a}$-dependent part of the path-integral, which we call
$I[h]$,
\begin{equation}
I[h]\equiv\int {\cal D}{{\cal H}_{\mu}}^{a}\exp\left(i\int
d^3x\left(\frac{1}{2}{{\cal
H}_{\mu}}^{a}{{[{S_a}^{\mu}]}_b}^{\rho}{{\cal H}_{\rho}}^{b} +
{{\cal H}_{\mu}}^{a}{J_a}^{\mu}\right)\right)
\end{equation}
where we have defined
\begin{eqnarray}
{{[{S_a}^{\mu}]}_b}^{\rho}&=&m\eta_{ab}\varepsilon^{\mu\nu\rho}\partial_{\nu},\label{s}\\
{J_a}^{\mu}&=&-m\varepsilon^{\mu\nu\lambda}\partial_{\nu}h_{\lambda
a}=-{{[{S_a}^{\mu}]}_b}^{\rho}{h_{\rho}}^b.
\end{eqnarray}
At this point it should be remarked that due to gauge
invariance operator
${{[{S_a}^{\mu}]}_b}^{\rho}$ is non-invertible. As usual, this problem can be overcome
by adding an appropriate gauge fixing term, this leading to  a ``regulated''
invertible operator (which we call ${{[{{(S^{reg})}_a}^{\mu}]}_b}^{\rho}$) so that
 $I[h]$ can be written as
\begin{equation}
I[h] =\int {\cal D}{{\cal H}_{\mu}}^{a} \exp\left(i\int
d^3x\left((\frac{1}{2}{{\cal
H}_{\mu}}^{a}{{[{{(S^{reg})}_a}^{\mu}]}_b}^{\rho} {{\cal
H}_{\rho}}^{b} + {{\cal H}_{\mu}}^{a}{J_a}^{\mu}\right)\right) \label{pi}
\end{equation}
As usual, at the end of the calculations the regulator can be turned off
and a finite result is attained.

Being quadratic in ${{\cal H}_\mu}^a$, path-integral (\ref{pi})
can be accommodated as
\begin{eqnarray}
I[h] &=&\int {\cal D}{{\cal H}_{\mu}}^{a} \exp\left(
\frac{i}{2}\int d^3x({{\cal
H}_{\mu}}^{a}-{h_{\mu}}^{a}){{[{{(S^{reg})}_a}^{\mu}]}_b}^{\rho}({{\cal
H}_{\rho}}^{b}-{h_{\rho}}^{b})
\right)   \nonumber\\
&\times&\exp\left(i\int
d^3x\left(-\frac{1}{2}{{[{{(S^{reg})}_a}^{\mu}]}_b}^{\rho}{h_{\rho}}^{b}{{[{{({S^{reg})}^{-1}}_{\mu}}^{a}]}_{\lambda}^{c}
{[{{(S^{reg})}_c}^{\lambda}]}_d}^{\nu}{h_{\nu}}^{d}\right) \right)
\label{formula}
\end{eqnarray}
After a shift ${{\cal H}_{\mu}}^a - {h_{\mu}}^a \to {{\cal
H}_{\mu}}^a$ in the path-integral variables the factor in the
first line of (\ref{formula}) becomes $h$-independent giving
a field independent constant factor ${\cal N}_1$ that will be irrelevant for the
calculation of vacuum expectation values from the generating functional.
We can then write eq.(\ref{formula}) in the form
\begin{equation}
I[h]= {\cal N}_1 \,\exp\left(-\frac{i}{2}\int
d^3x{{[{{(S^{reg})}_a}^{\mu}]}_b}^{\rho}{h_{\rho}}^{b}{{[{{({S^{reg})}^{-1}}_{\mu}}^{a}]}_{\lambda}^{c}
{[{{(S^{reg})}_c}^{\lambda}]}_d}^{\nu}{h_{\nu}}^{d} \right)
\end{equation}
One can now cancel out the $(S^{reg})^{-1} S^{reg}$ factor in the integral of the exponential factor
and then turn off the regulator so that one finally ends with
\begin{equation}
I[{h_{\mu}}^{a}]= {\cal N}_1 \exp\left(i\int
d^3x\left(-\frac{m}{2}{h_{\mu}}^{a}\varepsilon^{\mu\nu\lambda}\partial_{\nu}h_{\lambda
a}\right)\right)
\end{equation}
Putting all this together we have
\begin{equation}
Z_I[j]={\cal N}_1 \int{\cal D}{h_{\mu}}^{a}\exp\left(i({S}_{SD1}[h]
+i \int d^3x h_\mu^{\, \; a} j^{\;\, \mu}_{a})\right)
\label{ZIJfinal}
\end{equation}
or
\begin{equation}
Z_I[j] = {\cal N}_1 Z_{SD1}[j], \label{primera}
\end{equation}
where $Z_{SD1}[j]$ is the generating functional for a self-dual
spin two model, in the presence of a source, with classical action $S_{SD1}$ as defined in the
previous section, eq.(\ref{sd1action}).

We shall now start again from $Z_I[j]$ as given by (\ref{ZIJ}) but now we shall perform
 the ${h_{\mu}}^{a}$ integral. As before, we write the path integral  as
\begin{equation}
I[H]=\int {\cal D}{h_{\mu}}^{a}\exp\left(i\int
d^3x\left(\frac{1}{2}{h_{\mu}}^{a}{{[{D_a}^{\mu}]}_b}^{\rho}{h_{\rho}}^{b}
+ {h_{\mu}}^{a}{{\cal J}_a}^{\mu}\right)\right), \label{uia}
\end{equation}
where we have defined
\begin{eqnarray}
{{[{D_a}^{\mu}]}_b}^{\rho}&=&-m^2\varepsilon_{abc}\varepsilon^{\mu\rho\lambda}{\delta_{\lambda}}^{c},\\
{{\cal J}
_a}^{\mu}&=&-m\varepsilon^{\mu\nu\lambda}\partial_{\nu}{\cal
H}_{\lambda a} + j^{\;\, \mu}_{a}
=\frac{1}{m}{{[{D_a}^{\mu}]}_b}^{\rho}{W_{\rho}}^b({\cal H}) +
j^{\;\, \mu}_{a} \label{last}
\end{eqnarray}
and ${W_{\rho}}^b({\cal H})$ is the linear spin connection as
defined in (\ref{tres}) but for the dual field ${{\cal
H}_{\lambda}}^{\,\;a}$. The last equality in (\ref{last}) follows
from
 (\ref{con}).
  Now, we can complete squares in (\ref{uia}) getting
\begin{eqnarray}\label{iH}
&& I[{{\cal H}_{\mu}}^{a}]=\int {\cal
D}{h_{\mu}}^{a}\exp\left(i\int
d^3x\left(\frac{1}{2}\varepsilon^{\mu\nu\lambda}\partial_{\nu}{{\cal
H}_{\lambda}} ^a{W_{\mu}}^a({\cal H}) -\frac1m W_\mu^a[{\cal
H}]j^{\;\, \mu}_a -\frac{1}{2} j^{\;\, \mu}_a
{[{{D^{-1}}_{\mu}}^{a}]_{\nu}}^{b} j^{\, \; \nu}_b
\right. \right.\nonumber\\
&&~~\left. \left. +
\frac{1}{2}\left({h_{\mu}}^a+\frac{1}{m}{W_{\mu}}^a({\cal H}) +
{[{{D^{-1}}_{\mu}}^{a}]_{\nu}}^{b}j^{\, \; \nu}_b
\right){{[{D_a}^{\mu}]}_d}^{\rho}\left({h_{\rho}}^d+\frac{1}{m}{W_{\rho}}^d({\cal
H}) + {[{{D^{-1}}_{\rho}}^{d}]_{\sigma}}^{c}j^{\; \, \sigma}_c
\right) \right)\right)\nonumber\\
\end{eqnarray}
where ${[{{D^{-1}}_{\mu}}^{a}]_{\nu}}^{b}$ is the inverse of
${{[{D_a}^{\mu}]}_b}^{\rho}$,
\begin{equation}
{[{{D^{-1}}_{\mu}}^{a}]_{\nu}}^{b} = \frac{1}{m^2} \left(\frac12
\delta_\mu^{\;\,a}\delta_\nu^{\;\,b}
-\delta_\mu^{\;\,b}\delta_\nu^{\;\,a} \right) \label{terin}
\end{equation}
The sign of the first term in (\ref{iH}) is the correct sign for
the Einstein action and it does not depend of $m$, so it will be
unaffected if we change the signs of the first two terms in the
interpolating action. The gaussian path-integration
over ${h_{\mu}}^{a}$ can be easily performed through a shift in
${h_{\mu}}^a$ leading to a multiplicative irrelevant factor ${\cal N}_0$
so that finally we have for $Z_I$
\begin{equation}
Z_I[j] = {\cal N}_0\int{\cal D}{{\cal H}_{\mu}}^{a}\exp\left(i\left(
S_{SD2} - \int d^3x \left( \frac1m W_\mu^a[{\cal H}]j^\mu_a
+\frac12 j^{\, \;\mu}_a {[{{D^{-1}}_{\mu}}^{a}]_{\nu}}^{b} j^{\,
\;\nu}_b \right) \right) \right) \label{acople}
\end{equation}
Using eq.(\ref{terin}) one can see  that when computing
correlation functions, the term quadratic in $j$ in
eq.(\ref{acople}) will give contact terms  which as usual can be
handled introducing an appropriate regularization.

In conclusion, we have obtained now that the generating functional
$Z_I[j]$ is also equivalent to the one for the $S_{SD2}$ action
\begin{equation}
Z_I[j] = {\cal N}_0 Z_{SD2}[j_{{}_W}] \label{segunda}
\end{equation}
where the coupling of the external source $j$  to ${\cal H}$ in
$Z_I[j]$ is now that defined in (\ref{acople}),
\begin{equation}
Z_{SD2}[j_{{}_W}] = \int {\cal D}{\cal H} \exp\left(iS_{SD2}[{\cal
H},j_{{}_W}] \right) \label{SSS}
\end{equation}
\begin{equation}
 S_{SD2}[{\cal H},j_{{}_W}] =S_{SD2}[{\cal H}] - \int d^3x \left(
\frac1m W_\mu^a[{\cal H}]j^\mu_a +\frac12 j^{\, \;\mu}_a
{[{{D^{-1}}_{\mu}}^{a}]_{\nu}}^{b} j^{\, \;\nu}_b \right)
\label{SSSS}
\end{equation}
The subindex $W$ in $j_W$ is included to recall  that in generating functional $Z_{SD2}$ the source couples (non-minimally)
  to ${{\cal H}_{\mu}}^a$ through the connection
${W_{\mu}}^a({\cal H})$.

Now,  comparing the two results obtained by
integrating $h$ and $H$ in interchanged orders, eqs.(\ref{primera})
and (\ref{segunda}), we conclude that, up to a constant
multiplicative factor, the following identity holds for spin-2 generating functionals
\begin{equation}
 {\cal N}_1 Z_{SD1}[j] = {\cal N}_0 Z_{SD2}[j_{{}_W}]
 \label{56}
\end{equation}
Differentiating both sides of this equation with respect to the
source and then making $j=0$ we have the following identity for
vacuum expectation values
\begin{equation}
\langle h_\mu^a(x) \rangle_{SD1} = -\frac1m\left \langle
W_\mu^a[{\cal H}(x)] \right\rangle_{\!\!SD2}
\label{rita}
\end{equation}
 We see that we have established  a duality relation
which holds at the quantum level and which can be written, in terms of fields as
\begin{equation}
h_\mu^a(x)  \to -\frac1m W_\mu^a[{\cal H}(x)] \label{dualss}
\end{equation}
One can see that duality relation (\ref{dualss}) also holds when
calculating correlation functions of an arbitrary number of $h$
fields in the self-dual theory SD1 in the sense that the answer can
be analogously calculated from products of spin connections
$W_\mu^a[{\cal H}]$ in the SD2 theory
\begin{equation}\label{correlacion}
\langle {h_{{\mu}_1}}^{{a}_1}(x_1) \cdots
{h_{{\mu}_n}}^{{a}_n}(x_n) \rangle_{SD1} = \left \langle
{(-\frac1m)}^n {W_{{\mu}_1}}^{a_1}[{\cal H}(x_1)] \cdots
{W_{{\mu}_n}}^{a_n}[{\cal H}(x_n)] \right\rangle_{\!\!SD2} + {\rm
contact\; terms}.
\end{equation}
We also note that the identification is consistent with the form
of the $SD2$ action given in (\ref{sd2action2}). This fact will
appear again when we consider the connection with the
topologically massive model in the next section. Let us end this
section by noting that since the coupling term ${W_{\mu}}^a({\cal
H}){j_a}^{\mu}$ is invariant under diffeomorphism it does not
constraint the the external source.
\section{Connection with Topologically Massive Gravity (TMG)}
We can now make the connection between the intermediate model
(SD2)~\cite{ArKh} and the linear topologically massive (SD3) model
\cite{DJT}. For this purpose we start with a new interpolating
action
\begin{equation}
S'_I[{h_{\mu}}^{a},{{\cal H}_{\mu}}^{a}]=\frac{1}{2}\int
d^3x\left(m{h_{\mu}}^{a}\varepsilon^{\mu\nu\lambda}\partial_{\nu}h_{\lambda
a}-{H_{\mu\nu}}G^{\mu\nu}({\cal H})+\frac{1}{2}{{\cal
H}_{\mu\nu}}G^{\mu\nu}({\cal H})\right)
\label{calor}
\end{equation}
where
\be
H_{\mu\nu}=h_{\mu\nu}+h_{\nu\mu}
\label{simetrizacion}
\ee
 and ${\cal H}_{\mu\nu}$
is its (symmetric) dual field (only
the symmetric part of the dreibein deviation contributes in the
mixing term). Action (\ref{calor}) is invariant, up to a surface term, under the following
diffeomorphism transformations
\begin{equation}
\delta{h_{\mu}}^{a}=\partial_{\mu}\xi^a \; , \qquad
\qquad \delta{\cal
H}_{\mu\nu}=\partial_{\mu}\zeta_{\nu}+\partial_{\nu}\zeta_{\mu}
\end{equation}
which corresponds to diffeomorphism transformations. The terms
containig the field ${\cal H}$ are explicitly Lorentz invariant.

As in the previous cases,  we introduce a partition function associated to $S'_I$
\begin{equation}
Z'_I=\int{\cal D}{h_{\mu}}^{a}{\cal D}{{\cal
H}_{\mu\nu}}\exp(iS'_I[h,{\cal H}])
\end{equation}
and   the corresponding generating functional
\begin{equation}
Z'_I[j]=\int{\cal D}{h_{\mu}}^{a}{\cal D}{{\cal
H}_{\mu\nu}}\exp(iS'_I[h,{\cal H},j]) \label{zprima}
\end{equation}
with
\begin{equation}\label{sprimaj}
S'_I[h,{\cal H},j]=S'_I[h,{\cal H},]+\int d^3x
{h_{\mu}}^a{j_a}^{\mu}
\end{equation}
and $j_a^{\, \; \mu}$ the external source. This source should
satisfy $\partial_{\mu}{j_a}^{\mu}=0$  in order to preserve
invariance   under diffemorphisms.

To connect $Z'_I[j]$ with $Z_{SD2}[j]$, the generating functional
of the SD2 theory defined in eq.(\ref{sd2action2}), we shall  perform
the ${\cal H}_{\mu\nu}$-integration in (\ref{zprima})
\begin{equation}\label{ih}
I[{h_{\mu}}^{a}]=\int {\cal D}{{\cal H}_{\mu\nu}}\exp\left(i\int
d^3x\left(\frac{1}{4}{{\cal H}_{\mu\nu}}{{[{{\cal
G}}^{\mu\nu}]}_{\lambda\rho}}{{\cal H}^{\lambda\rho}} + {{\cal
H}_{\mu\nu}}{{J}}^{\mu\nu}\right)\right)
\end{equation}
where we have defined now
\begin{eqnarray}
&& {{[{{\cal
G}}^{\mu\nu}]}_{\lambda\rho}}{{\cal H}^{\lambda\rho}} = G^{\mu\nu}({\cal H})\\
&& {J}^{\mu\nu} = -\frac{1}{2}{{[{{\cal
G}}^{\mu\nu}]}_{\lambda\rho}}{H^{\lambda\rho}}
\end{eqnarray}
The operator $-(1/2){{[{{\cal G}}^{\mu\nu}]}_{\lambda\rho}}$, the
evolution operator of the linear Einstein action, is
non-invertible due to the gauge invariance and should be
regularized. After integration is performed the regularization can be
turned off as in the previous case (see
eqs.(\ref{ZIJ})-(\ref{ZIJfinal})). The final  answer is
\begin{equation}
I[{h_{\mu}}^{a}]=\exp\left(-\frac{i}{4}\int d^3x{H_{\mu\nu}}G^
{\mu\nu}(H)\right)\int {\cal D}{{\cal H}_{\mu\nu}}\exp\left(
\frac{i}{4}\int d^3 x({{\cal H}_{\mu\nu}}-{H_{\mu\nu}}){[{{\cal
G}}^{\mu\nu}]}_{\lambda\rho}({{\cal
H}^{\lambda\rho}}-{H^{\lambda\rho}})\right) \label{pis}
\end{equation}
Making  the shift  ${{\cal H}_{\mu\nu}}-{H_{\mu\nu}} \to {{\cal
H}_{\mu\nu}}$ any dependence on $H_{\mu\nu}$ in the path-integral factor   disappears. Calling  ${\cal N}_2$ this factor,
    irrelevant when computing
vacuum expectation values, we end with
\begin{equation}
I[{h_{\mu}}^{a}]={\cal N}_2 \exp\left(-\frac{i}{4}\int
d^3x{H_{\mu\nu}}G^ {\mu\nu}(H)\right)
\end{equation} Inserting
this result in expression (\ref{zprima}) for $Z'_I[j]$, we see
that the partition function for the SD2 theory (with minimal
coupling ${h_{\mu}}^a\!{j_a}^{\mu}$) takes the form
\begin{equation}
Z'_I[j]={\cal N}_2\int {\cal D}{h_{\mu}}^{a}\exp[iS_{SD2}[h,j]] =
{\cal N}_2Z_{SD2}[j] \label{65}
\end{equation}

As in the previous section, we now invert the integration order in (\ref{zprima})
starting from the ${h_{\mu}}^{a}$ path-integral,
\begin{equation}
I'[{\cal H}]=\int {\cal D}{h_{\mu}}^{a}\exp\left(i\int
d^3x\left(\frac{1}{2}{h_{\mu}}^{a}{{[{S_a}^{\mu}]}_b}^{\rho}{h_{\rho}}^{b}
+ {h_{\mu}}^{a}{{\cal J}_a}^{\mu}\right)\right)\label{sj}
\end{equation}
where ${{[{S_a}^{\mu}]}_b}^{\rho}$ is defined in eq.(\ref{s}) and
\begin{equation}
{{\cal J}_a}^{\mu}=j_a^{\, \; \mu}
+\frac{1}{m}{{[{S_a}^{\mu}]}_b}^{\rho}{W_{\rho}}^b({{\cal H}})
\end{equation}
Following the same procedure as for the previous
calculation, the integration can be made straightforwardly after completing squares and introducing an appropriate regularization of
${{[{S_a}^{\mu}]}_b}^{\rho}$.  The answer is
\begin{equation}
I'[{\cal H}]={\cal N}_1 \exp\left(i\int
d^3x\left(-\frac{1}{m}{W_{\mu}}^a({\cal
H})\varepsilon^{\mu\nu\lambda}\partial_{\nu}W_{\lambda a}({\cal
H})-\frac{1}{m}{W_{\mu}}^a({\cal
H}){j_a}^{\mu}-\frac{1}{2}{j_a}^{\mu}
{{[{S^{-1}_{\mu}}^{a}]}_{\nu}}^{b}{j_b}^{\nu}\right) \right)
\end{equation}
where as before we have factorized a field-independent irrelevant
constant ${\cal N}_1$ arising from the quadratic
 path-integral.
We then have for $Z'_I[j]$
\begin{equation}
Z'_I[j]={\cal N}_1\int {\cal D}{{\cal
H}_{\mu}}^{a}\exp\left(iS_{SD3}[{\cal H},j_{{}_W}]\right) = {\cal
N}_1Z_{SD3}[j_{{}_W}] \label{69}
\end{equation}
where we have defined
\begin{equation}
Z_{SD3}[j_{{}_W}] = \int {\cal D}{\cal H}_\mu^{\,a}
\exp\left(iS_{SD3}[{\cal H},j_{{}_W}] \right)
\end{equation}
with
\begin{equation}
S_{SD3}[{\cal H},j_{{}_W}] = S_{SD3}[{\cal H}] - \int d^3x \left(
\frac{1}{m}{W_{\mu}}^a({\cal H}){j_a}^{\mu}+\frac{1}{2}{j_a}^{\mu}
{{[{S^{-1}_{\mu}}^{a}]}_{\nu}}^{b}{j_b}^{\nu} \right)
\label{sietecinco}
\end{equation}
and $S_{SD3}[{\cal H}]$ defined in eq.(\ref{sd3action2}). It is
worth noting at this point that in general if we couple the
connection with a conserved source the only contribution will come
from the part in $W$ that depends on the symmetric part of the
dreibein deviation. In fact from (\ref{seis})
\begin{equation}
\partial_{\mu}{j_a}^{\mu}=0 \; \to \;{W_{\mu}}^a(h){j_a}^{\mu}={W_{\mu}}^a(H){j_a}^{\mu}+ {\rm \,surface\;term}
\end{equation}
so this coupling term will be explicitly Lorentz invariant if we
work with a general dual field ${\cal H}$.

 Comparing (\ref{65}) and
(\ref{69}) we obtain the main result in this section
\begin{eqnarray}
{\cal N}_2Z_{SD2}[j] = {\cal N}_1Z_{SD3}[j_{{}_W}] . \label{76}
\end{eqnarray}
We again stress that the Einstein term in $SD3$ appears with a minus sign in front as
it should be in order to have the correct description of a $+2$
helicity excitation with mass $m$. As we said before this cannot
be changed modifying $S'_I$ because we will loose the connection
with $SD2$.

Differentiation with respect to the source leads to the duality
connection between
 vacuum expectation values
\begin{equation}
\langle h_\mu^a \rangle_{SD2} = - \left \langle \frac1m
W_\mu^a[{\cal H}] \right\rangle_{\!\!SD3}
\label{mediomedio}
\end{equation}
We have again established
  a duality relation
which holds at the quantum level, this time between theories SD2 and SD3.

This identification at the quantum level is consistent with the classical duality result. Indeed, considering (\ref{seis}) and (\ref{diez}) in
(\ref{sd3action2})
\begin{eqnarray}
S_{SD3}&=&\frac{1}{2}\int d^{3}x\left({h_{\mu}}^{a}{{{{K^{\pm}}_a}^
{\mu}}_{b}}^{\lambda}\left(-\frac{1}{m^2}{S_{\lambda}}^b(h)\right)\right)\nonumber
\\
&=&\frac{1}{2}\int d^{3}x\left({h_{\mu}}^{a}{{{{K}_a}^
{\mu}}_{b}}^{\lambda}\left(-\frac{1}{m}{[{W_{\lambda}}^b]_c}^{\sigma}\right)\left(-\frac{1}{m}{W_{\sigma}}^c(h)\right)\right)\\
&=&\frac{1}{2}\int d^{3}x\left({h_{\mu}}^{a}{{{{K^{SD2}}_a}^
{\mu}}_{b}}^{\lambda}\left(-\frac{1}{m}{W_{\lambda}}^b(h)\right)\right),
\end{eqnarray}
where we have identified the evolution operator of the $SD2$ model
with ${{{{K^{SD2}}_a}^ {\mu}}_{b}}^{\lambda}$.

Concerning correlation functions, in this case the presence of the
source terms in (\ref{sietecinco}) should be appropriately
handled. The conserved source can be decomposed as
\begin{equation}\label{jotass}
{{j_a}}^{\mu}=({{j_a}}^{0},{{j_a}}^{i})=(J_a,\varepsilon_{ij}\partial_j
J^T_a + \partial_i(-\Delta)^{-1}{\dot J}_a),
\end{equation}
and it can be seen that for conserved sources the quadratic term
in (\ref{sietecinco}) can be written as
\be\label{ochoseis}
\frac{1}{2}\int d^3 x{{j_a}}^{\mu}
{{[{S^{-1}_{\mu}}^{a}]}_{\nu}}^{b}{{j_b}}^{\nu}=\frac{1}{2m}\int
d^3
x{{j_a}}^{\mu}{\eta}^{ab}\varepsilon_{\mu\rho\nu}\frac{\partial^{\rho}}{\Box}{{j_b}}^{\nu}=
 \frac{1}{m}\int d^3 x {\eta}^{ab}J_aJ^T_b
\ee
In this sense we
get a result equivalent to (\ref{correlacion}), now relating  correlation
functions for $SD2$ and $SD3$,
\begin{equation}\label{correlacion23}
\langle {h_{{\mu}_1}}^{{a}_1}(x_1) \cdots
{h_{{\mu}_n}}^{{a}_n}(x_n) \rangle_{SD2} = {(-\frac1m)}^n \left
\langle
 {W_{{\mu}_1}}^{a_1}[{\cal H}(x_1)] \cdots
{W_{{\mu}_n}}^{a_n}[{\cal H}(x_n)] \right\rangle_{\!\!SD3} + \; \;
{\rm contact\; terms}
\end{equation}
so that a field mapping reproducing the quantum relations (\ref{mediomedio})-(\ref{correlacion23})  can be also written  for the SD2 $\to$ SD3
duality relation
\be
h_\mu^a(x) \rightarrow - \frac1m
W_\mu^a[{\cal H}](x)
\ee

Let us end by noting that the coupling term $({1}/{m}){W_{\mu}}^a{j_a}^{\mu}$ in (\ref{sietecinco}) is invariant
under diffeomorphism and, as we pointed, under Lorentz
transformations due to the transversality of the source. So the
map preserves the gauge invariances without imposing new
constraints on the source.

We can extend the connection to the self-dual model SD1 taking
into account (\ref{acople}) and changing the source coupling in
$S'_I$ in (\ref{sprimaj}). To this end,  instead of adding a term of the form
${h_{\mu}}^a{j_a}^{\mu}$, we add a source coupling term of the form
$-({1}/{m}){W_{\mu}}^a(h){j_a}^{\mu}$, which also preserves the
gauge invariance. We start the process with a new generating
interpolating function ${\widetilde{Z'}}_I[j_{{}_W}]$ and after integrating  over ${{\cal H}_{\mu\nu}}$ gets  the $SD2$
model with the appropriate coupling
\begin{eqnarray}
{\widetilde{Z'}}_I[j_{{}_W}]&=&\int{\cal D}{h_{\mu}}^{a}{\cal
D}{{\cal H}_{\mu\nu}}\exp(i{\widetilde{S'}}_I[h,{\cal H},j_W]), \nonumber \\
&=&\int{\cal D}{h_{\mu}}^{a}{\cal
D}{{\cal H}_{\mu\nu}}\exp\left(i(S'_I[h,{\cal H}]-\int d^3x\frac{1}{m} {W_{\mu}}^a(h){j_a}^{\mu})\right) \nonumber \\
&=&{\cal N}_2\int{\cal D}{h_{\mu}}^{a}\exp\left(i(S_{SD2}[h]-\int
d^3x\frac{1}{m} {W_{\mu}}^a(h){j_a}^{\mu})\right)\\
&=& {\cal N}_2{\widetilde{Z}}_{SD2}[j_{{}_W}].
\end{eqnarray}

Performing instead the path-integration over ${h_{\mu}}^a$, we get an expression
like (\ref{sj}) but where  ${{\cal{J}}_a}^{\mu}$ now reads
\begin{equation}
{{\cal{J}}_a}^{\mu}=-\frac{1}{m^2}{{[{S_a}^{\mu}]}_b}^{\rho}({{\widetilde{j}}_{\rho}}^b-m{W_{\rho}}^b({\cal
H})),
\end{equation}
with
\begin{equation}\label{jtilde}
{{\widetilde{j}}_{\rho}}^b=({\delta_{\mu}}^b{\delta_{\rho}}^a-\frac{1}{2}{\delta_{\mu}}^a{\delta_{\rho}}^b){j_a}^{\mu}.
\end{equation}
Then, after the integration over  ${h_{\mu}}^a$ we obtain
\begin{eqnarray}
{\widetilde{Z}}'_I[j_{{}_W}]&=&{\cal N}_1\int {\cal D}{{\cal
H}_{\mu}}^{a}\exp[i(S_{SD3}[{\cal H}]-\int
d^3x(\frac{1}{2m^3}{{\widetilde{j}}_{\mu}}^a\varepsilon^{\mu\nu\lambda}\partial_{\nu}{\widetilde{j}}_{\lambda
a}+\frac{1}{m^2}{{\widetilde{j}}_{\mu}}^a{G^{\mu}}_a[{\cal H}]
))],  \nonumber \\
&=&{\cal N}_1\int {\cal D}{{\cal H}_{\mu}}^{a}\exp[i(S_{SD3}[{\cal
H}]-\int
d^3x(\frac{1}{2m^3}{{\widetilde{j}}_{\mu}}^a\varepsilon^{\mu\nu\lambda}\partial_{\nu}{\widetilde{j}}_{\lambda
a}+\frac{1}{m^2}{{j}_a}^{\mu}{S^{\mu}}_a[{\cal H}]))]\label{sd31}
\end{eqnarray}
where we have denoted  the direct coupling term in the form
${{\widetilde{j}}_{\mu}}^a{G^{\mu}}_a({\cal H})={S_{\mu}}^a({\cal
H}){j_a}^{\mu}$, with ${S_{\mu}}^a$ the Schouten tensor. We
conclude that
\begin{eqnarray}
{\cal N}_2 {\widetilde{Z}}_{SD2}[j_{{}_W}] = {\cal
N}_1{\widetilde{Z}}_{SD3}[j_{{}_S}]
\end{eqnarray}
where the subindex $S$ in the source indicates that the coupling
in $SD3$ is that written in eq. (\ref{sd31}). The coupling term
with the external source is explicitly Lorentz invariant due to
the fact that the Schouten tensor depends on the symmetrical part
of ${{\cal H}_{\mu\nu}}$, so as before the gauge invariances of
the action are regained by the mapping.

The symmetry of the Schouten tensor implies that only the symmetric part of the source
$j^s_{\mu\nu}=(1/2)(j_{\mu\nu}+j_{\nu\mu})$ contributes to the coupling
so the connection
between correlation functions should be    obtained by differentiating with respect
to $j^s_{\mu\nu}$.  Taking this into account  the following relation between vacuum expectation values hold
\begin{equation}\label{vev123}
\frac{1}{2} \left\langle {H_{\mu\nu}}(x) \right\rangle_{SD1} = -\frac{1}{m}\langle
 {W^s_{\mu\nu}}(H)(x) \rangle_{SD2} = - \frac{1}{m^2}\left \langle
{S_{\mu\nu}} (H)(x) \right\rangle_{SD3}
\end{equation}
where the supper script $s$ indicates that symmetrization is assumed. Note that solely $H_{\mu\nu}$,
 the symmetric part of $h_{\mu\nu}$, appears in (\ref{vev123}) due  to the conservation condition imposed to
 the source. The result should be compared with  (\ref{rita}) where the complete $h_{\mu\nu}$  tensor
 contributes.

The relation between the vacuum expectation values that we have obtained  is consistent
with the form of the classical actions in (\ref{sd2action2}) and
(\ref{sd3action2}). We note that in each mapping the v.e.v. of $h$ is
mapped in a v.e.v. of  gauge invariant objects of each  model.

\section{Connection with the new topologically massive gravity}
As before, we shall work in terms of the deviation
of the metric (see (\ref{sd3actionsim}) and
(\ref{sd4action})). The interpolating action with minimal coupling
is now
\begin{equation}
S''_I[H,{\cal H},j]=\frac{1}{4}\int d^3 x\left(-\frac{1}{m}{\cal
H}_{\mu\nu}C^{\mu\nu}({\cal H})+\frac{2}{m}{\cal
H}_{\mu\nu}C^{\mu\nu}(H)+H_{\mu\nu}G^{\mu\nu}(H)+2H_{\mu\nu}j^{\mu\nu}\right),
\label{sint34}
\end{equation}
where $H_{\mu\nu}$ is the deviation of the metric and ${\cal
H}_{\mu\nu}$ the dual field, which is assumed to
be symmetric; $j^{\mu\nu}$ is a symmetric conserved source.
Action (\ref{sint34}) is invariant, up to surface terms, under diffeomorphism
transformations in the two fields and under conformal
transformations in the dual field
\begin{equation}
\delta H_{\mu\nu}=\partial_{\mu}\xi_{\nu}+\partial_{\nu}\xi_{\mu}
\; \; ; \; \; \delta{\cal H}_{\mu\nu}=\partial_{\mu}{\tilde {
\xi}}_{\nu}+\partial_{\nu}{\tilde{\xi}}_{\mu}+\eta_{\mu\nu}\rho
\end{equation}

We  now introduce the generating functional
\begin{equation}
Z´´_I[j]=\int {\cal D}H_{\mu\nu}{\cal D}{\cal
H}_{\mu\nu}\exp\left(iS''_I[H,{\cal H},j]\right)
\end{equation}
and, as in the previous sections we shall integrate in the two opposite orders.
First we shall consider the integral over the dual field ${\cal
H}$, this allowing to  connect the generating functional
$Z´´_I[j]$  with that of the linear topologically massive model
$SD3$ .   We start defining the functional
\begin{equation}
I[H_{\mu\nu}]=\int {\cal D}{\cal H}_{\mu\nu}\exp\left(\frac{i}{4}\int
d^3 x\left(-\frac{1}{m}{\cal H}_{\mu\nu}{\cal
C}^{\mu\nu,\lambda\rho}{\cal H}_{\lambda\rho}+\frac{2}{m}{\cal
H}_{\mu\nu}J^{\mu\nu}\right)\right)
\end{equation}
where  operator ${\cal
C}^{\mu\nu,\lambda\rho}$,   acting on ${\cal
H}_{\mu\nu}$ gives the linear Cotton tensor
\begin{equation}
{\cal C}^{\mu\nu,\lambda\rho}{\cal
H}_{\lambda\rho}={C^L}^{\mu\nu}({\cal H})
\end{equation}
and
\begin{equation}
J^{\mu\nu}\equiv {\cal C}^{\mu\nu,\lambda\rho}H_{\lambda\rho}.
\end{equation}

As it stands, operator ${\cal C}^{\mu\nu,\lambda\rho}$ is non-invertible due
to gauge invariance, so it must be regularized. As before, we shall adopt some regularization which can be safely
  turned off at the end of the calculation. The answer is
\begin{eqnarray}
I[H_{\mu\nu}]&=&\exp\left(i\frac{1}{4m}\int d^3x
H_{\mu\nu}C^{\mu\nu}(H)\right)\times \nonumber
\\ &&\int {\cal D}{\cal
H}_{\mu\nu}\exp\left(-\frac{i}{4m}\int d^3 x({\cal
H}_{\mu\nu}-H_{\mu\nu}){\cal C}^{\mu\nu,\lambda\rho}({\cal
H}_{\lambda\rho}-H_{\lambda\rho})\right)\\
&=&{\cal N}_3 \; \exp\left(\frac{i}{4m}\int d^3x
H_{\mu\nu}C^{\mu\nu}(H)\right),
\end{eqnarray}
with ${\cal N}_3$ an irrelevant constant factor.

From the calculation above,   a connection with the generating functional of the
linear topologically massive model $SD3$ is established
\begin{equation}
Z''_I[j]={\cal N}_3Z_{SD3}[j] \label{bongo}
\end{equation}

We shall now follow the inverse  road, first integrating in $H_{\mu\nu}$, this
 allowing   to   connect $Z''_I[j]$ with the partition function of the   new topologically
massive model . The integral to be performed  can be written as
\begin{equation}\label{Ih4}
I[{\cal H}_{\mu\nu}]=\int  {\cal D}H_{\mu\nu} \exp\left(\frac{i}{2}\int
d^3 x \left(\frac{1}{2}H_{\mu\nu}[{\cal G}^{\mu\nu}]^{\lambda\rho}
H_{\lambda\rho}+ H_{\mu\nu}{\cal J}^{\mu\nu}\right)\right)
\end{equation}
with $-(1/2)[{\cal G}^{\mu\nu}]^{\lambda\rho}$ the evolution
operator of the linear Einstein term and
\begin{equation}
{\cal J}^{\mu\nu}=\frac{1}{2}j^{\mu\nu}+\frac{1}{4m}[{\cal
G}^{\mu\nu}]^{\lambda\rho}({\varepsilon_{\lambda}}^{\alpha\beta}\partial_{\beta}{\cal
H}_{\beta\rho}+{\varepsilon_{\rho}}^{\alpha\beta}\partial_{\beta}{\cal
H}_{\beta\lambda}). \label{J}
\end{equation}
To obtain this last equation we have used the identity
\begin{equation}
{\cal H}_{\mu\nu}C^{\mu\nu}(H)=\frac{1}{2}H_{\mu\nu}[{\cal
G}^{\mu\nu}]^{\lambda\rho}({\varepsilon_{\lambda}}^{\alpha\beta}\partial_{\beta}{\cal
H}_{\beta\rho}+{\varepsilon_{\rho}}^{\alpha\beta}\partial_{\beta}{\cal
H}_{\beta\lambda})+ {\rm  \; surface \; term}
\end{equation}

The operator $[{\cal G}^{\mu\nu}]^{\lambda\rho}$ requires a
regularization in order to become invertible. Once this is done
one can safely integrate and then turn off the regulator. The
answer is
\begin{eqnarray}
I[{\cal H}_{\mu\nu}]&=&{\cal N}_2 \times \exp\left(-i\int d^3
x\left(\frac{1}{4m^2}{\varepsilon_{\mu}}^{\alpha\beta}\partial_{\alpha}{\cal
H}_{\beta\nu}[{\cal
G}^{\mu\nu}]^{\lambda\rho}{\varepsilon_{\lambda}}^{\gamma\sigma}\partial_{\gamma}{\cal
H}_{\sigma\rho} \right.\right.\nonumber \\
&+& \left.\left.
\frac{1}{2m}{\varepsilon_{\lambda}}^{\alpha\beta}\partial_{\alpha}{\cal
H}_{\beta\rho}j^{\lambda\rho}+\frac{1}{4}j^{\mu\nu}[{\cal
G}^{-1}_{\mu\nu}]_{\lambda\rho}j^{\lambda\rho}\right)\right).
\end{eqnarray}
The direct coupling term between the source and ${\cal
H}_{\mu\nu}$ is just $-({1}/{m})W_{\mu\nu}({\cal H})j^{\mu\nu}$.
Due to the symmetry of the source it is explicitly conformal
invariant since it depends solely on the symmetric part of the
linear connection. Hence, conformal invariance is regained in the map
without imposing new constraints on the source.

After these calculations we can then write
\begin{equation}
Z''_I[j] ={\cal N}_2Z_{SD4}[j_{{}_W}]\label{zetas4}
\end{equation}
with
\begin{equation}\label{extraj}
S_{SD4}[j_{{}_W}]=S_{SD4}-\int d^3 x
\left(\frac{1}{m}W_{\lambda\rho}({\cal
H})j^{\lambda\rho}+\frac{1}{4}j^{\mu\nu}[{\cal
G}^{-1}_{\mu\nu}]_{\lambda\rho}j^{\lambda\rho}\right)
\end{equation}

Then, from eqs. (\ref{bongo}) and (\ref{zetas4}) we finally have
the duality identity
\be {\cal N}_3Z_{SD3}[j] = {\cal
N}_2Z_{SD4}[j_{{}_W}]
 \ee
 which allows to write the following identity between vacuum expectation values in the two models
\begin{equation}
\frac{1}{2}\langle H_{\mu\nu}(x) \rangle_{SD3} = -\frac{1}{m}\left \langle
 W^s_{\mu\nu}[{\cal H} (x)]\right \rangle_{SD4}
\end{equation}
As in the previous sections this relation  is consistent with the
classical results. Indeed, from eq.(\ref{diez}) we can write the $SD4$ action in the form
\begin{eqnarray}
S_{SD4}&=&\frac{1}{4}\int d^3 x \left(H_{\mu\alpha}{K^
{\mu\alpha}}_{\rho\beta}\left(-\frac{1}{m^3}\varepsilon^{\rho\sigma\lambda}\partial_{\sigma}{{[W_{\lambda}}^{\beta}]_c}^{\gamma}{W_{\gamma}}^c(H)
\right)
\right)\nonumber
\\
&=&\frac{1}{4}\int d^3 x \left(H_{\mu\alpha}{K^
{\mu\alpha}}_{\rho\beta}\left(\frac{1}{m^3}{G^{\rho\beta}}W(H)\right)\right)
\nonumber \\
&=&\frac{1}{4}\int d^3 x \left(H_{\mu\alpha}{K^
{\mu\alpha}}_{\rho\beta}\left(-\frac{1}{m^2}{{\cal
S}^{\rho\beta}}_{\nu\gamma}\right)\left(-\frac{1}{m}\varepsilon^{\nu\sigma\lambda}\partial_{\sigma}{H_{\lambda}}^{\gamma}\right)\right)
\nonumber\\
&=&\frac{1}{4}\int d^3 x \left( H_{\mu\alpha}{{K^{SD3}}^
{\mu\alpha}}_{\rho\beta}\left(-\frac{1}{m}\varepsilon^{\rho\sigma\lambda}\partial_{\sigma}{H_{\lambda}}^{\beta}\right)\right)
\end{eqnarray}
 where  ${{\cal
S}^{\rho\beta}}_{\nu\gamma}$ as the differential operator that
acting on $H^{\nu\gamma}$ gives the linear Schouten tensor and
${{K^{SD3}}^ {\mu\alpha}}_{\rho\beta}$ the evolution operator associated to
action $SD3$.
Written the classical action in this form, the relation between classical and quantum relations becomes clear.

Concerning correlation functions of products of
fields one has to take care of contact terms resulting from the
 term quadratic in the source (see eq.(\ref{extraj})).   So also in this case we shall write
\begin{equation}\label{correlacion34}
\frac1{2^n} \langle ( H_{{\mu}_1{{\nu}_1}}(x_1)\! \cdots\!
H_{{\mu}_n{{\nu}_n}}(x_n) \rangle_{SD3} = {(-\frac{1}{m})}^n\left
\langle
 W^s_{{\mu}_1{{\nu}_1}}[{\cal H}(x_1)]\! \cdots\!
W^s_{{\mu}_n{{\nu}_n}}[{\cal H}(x_n)] \right\rangle_{\!\!SD4} +
{\rm \, contact\; terms}
\end{equation}
where the superscript $s$  indicates that the connection $W$  must be symmetrized.

Connections with
generating functionals of the other models can be established by changing   couplings to the source in
the interpolating action. In particular, taking into account
(\ref{acople}) one can change the minimal coupling term in
(\ref{sint34}) to the following one
\begin{equation}
-\frac{1}{2m^2}S_{\mu\nu}(H)j^{\mu\nu}=-\frac{1}{2m^2}H_{\mu\nu}{\cal
G}^{\mu\nu,\lambda\rho}{\widetilde{j}}_{\lambda\rho} + {\rm surface
\; term}
\end{equation}
with
${{\widetilde{j}}_{\rho}}^b=({\delta_{\mu}}^b{\delta_{\rho}}^a-\frac{1}{2}{\delta_{\mu}}^a{\delta_{\rho}}^b){j_a}^{\mu}$
as in (\ref{jtilde}).
If we now perform the integration over ${\cal H}$ we get the linear
topologically massive generating functional with the appropriate
coupling
\begin{equation}
Z''_I[j_{{}_S}]={\cal N}_3{\widetilde{Z}}_{SD3}[j_{{}_S}]
\end{equation}
The integration over $H_{\mu\nu}$ is analogous to (\ref{Ih4}) with
\begin{equation}
{\cal J}^{\mu\nu}=\frac{1}{2m}[{\cal
G}^{\mu\nu}]^{\lambda\rho}\left({\varepsilon_{\lambda}}^{\alpha\beta}\partial_{\beta}{\cal
H}_{\beta\rho}-\frac{1}{m}{\widetilde{j}}_{\lambda\rho}\right).
\label{J2}
\end{equation}
and leads to a generating functional in which there is a
non-minimal coupling to  the Cotton tensor
\begin{equation}
Z''_I[j_{{}_S}]={\cal N}_2 {\widetilde{Z}}_{SD4}[j_{{}_C}]
\end{equation}
with
\begin{equation}
S_{SD4}[j_{{}_C}]=S_{SD4}[{\cal H}]+\frac{1}{2}\int d^3
x\left(\frac{1}{m^3}C_{\mu\nu}({\cal
H})j^{\mu\nu}-\frac{1}{2m^2}{\widetilde {j}}_{\mu\nu}{\cal
G}^{\mu\nu,\lambda\rho}{\widetilde {j}}_{\lambda\rho}\right).
\end{equation}
This coupling is consistent with the classical expression of
action (\ref{sd4action2}) and it is clearly conformal invariant.

We can now make the correspondence of the v.e.v. from SD1 to
$SD4$, completing (\ref{vev123})
\begin{equation}\label{vev1234}
\frac{1}{2} \langle {H_{\mu\nu}}(x) \rangle_{SD1} = - \frac{1}{m} \langle
 {W^s_{\mu\nu}}(H)(x) \rangle_{SD2} = -\frac{1}{m^2} \left \langle
{S_{\mu\nu}} (H)(x) \right\rangle_{\!\!SD3}=\frac{1}{m^3}\left
\langle {C_{\mu\nu}} (H)(x) \right\rangle_{SD4}
\end{equation}

The connection with the SD2 could be obtained straightforward if
we start the process with a nonminimall coupling in (\ref{sint34})
with the symmetrized connection. In this case the integration in
${\cal H}_{\mu\nu}$ will give the SD3 generating function with the
appropriate coupling and the corresponding field independent
factor. The integration in $H_{\mu\nu}$ will take us to the
generating function of the SD4 model with a coupling with
$-\frac{1}{m^2}S_{\mu\nu}(H)$ as we would expect from the
classical description. The correspondence of the v.e.v. will be
\begin{equation}
\frac{1}{2} \langle {H_{\mu\nu}}(x) \rangle_{SD2} = -\frac{1}{m}\langle
 {W^s_{\mu\nu}}(H)(x) \rangle_{SD3} = -\frac{1}{m^2}\left \langle
{S_{\mu\nu}} (H)(x) \right\rangle_{SD4}.
\end{equation}

\section{Conclusions}
We have discussed the duality between four self-dual linearized
formulations for massive gravity in $2+1$ dimensions at the quantum level.
To this end, we have established generating functional connections in a
path-integral framework, this allowing  to find identities for vacuum
expectations values which  are consistent with the gauge
invariance of the classical actions.

We have found that the  v.e.v. for
the field ${h_{\mu}}^a$, the linear deviation of the dreibein in the self-dual model SD1, is mapped
(appart from a constant factor)  onto a the v.e.v  of the linear
connection $-({1}/{m}){W_{\mu}}^a(h)$ in the intermediate model
SD2, which is invariant under diffeomophisms. In turn, this
v.e.v. is mapped onto the vacuum expectation value  the Schouten
tensor $-({1}/{m^2}){S_{\mu}}^a(h)$ in the topologically massive
model SD3, which is invariant under diffeomorphism
transformations and explicitly under Lorentz transformations as it
depends on the symmetric part of the dreibein.   Finally, this
last v.e.v. is mapped into the v.e.v. of  the Cotton tensor,
$({1}/{m^3}){C_{\mu}}^a(h)$ which is clearly invariant under
diffeomorphism and conformal transformations. These
correspondences are consistent with the classical equations of
motion and  ensure the non-propagation of the lower spin parts of
${h_{\mu}}^a$ so that up to contact terms the mapping connects
 the spin 2 components of each object.

Following the same route, we have also connected the intermediate model
with the linear topologically massive and new topologically
massive models. In this case the ${h_{\mu}}^a$ coupling term is
invariant, up to a surface term, under diffeomorphisms if the source
is conserved ($\partial_{\mu}{j_a}^{\mu}=0$) and so is the
non-minimal coupling terms in the $SD3$ and $SD4$ models after the
dual map. The v.e.v. of ${h_{\mu}}^a$ is mapped first onto
$-({1}/{m}){W_{\mu}}^a(H)$ and then onto
$-({1}/{m^2}){S_{\mu}}^a(h)$.

The two linear topologically massive models are also connected and
again, after the dual map,  the coupling term  is invariant under
conformal transformations. The v.e.v. of $H_{\mu\nu}$ is mapped
onto $-({1}/{m}){W^s_{\mu\nu}}(H)$ .

The quantum identities that we have obtained  are consistent
with those one would expect from classical actions, as can be seen writing them in terms of
 the evolution operator of the self-dual model SD1,
as in (\ref{sd2action2}), (\ref{sd3action2}) and
(\ref{sd4action2}). The dual theories differ in their gauge
invariances and the couplings obtained when passing from one  to
the other are consistent.

The results described above can be schematized as follows

{
\footnotesize
\[
\begin{array}{cccc}
 {\rm  \underline{Self-dual~model}}\rightarrow & {\rm
\underline{2^{nd}~order~intermediate~model}}\rightarrow &   {\rm
\underline{TMG}}\rightarrow   &  \underline{\rm New~TMG} \cr ~
\vspace{-4mm} \cr S_{SD1}= \frac{1}{2}  \int hKh &
S_{SD2}=-\frac{1}{2m}\int  hKW[h]   &
\!\!\!S_{SD3}=\!-\frac{1}{4m^2}\int HK S[H] &
S_{SD4}=\frac{1}{4m^3}\int HK C[H]\cr ~ \vspace{-4mm} \cr \langle
{h_\mu}^a \rangle_{SD1} \rightarrow &   -\frac1m \langle
 W_\mu^a[h]  \rangle_{SD2} \rightarrow & -
 \frac{1}{m^2} \langle S_\mu^a[H]  \rangle_{SD3} & ~~~
 \frac{1}{m^3} \langle C_\mu^a[H]  \rangle_{SD4}\cr ~
\vspace{-4mm} \cr
 &\langle
h_\mu^a \rangle_{SD2} \rightarrow &  -  \frac1m\langle
W_\mu^a[H]  \rangle_{SD3} & - \frac{1}{m^2} \langle
S_\mu^a[H] \rangle_{SD4}\cr ~ \vspace{-4mm} \cr
 & &
\frac{1}{2} \langle H_{\mu\nu} \rangle_{SD3}  \rightarrow &- \frac{1}{m}  \langle   W^s_{\mu\nu}[{\cal H}
]  \rangle_{SD4} \cr
\end{array}
\]
}

\noindent with $K,W,S$ and $C$ defined in
 eqs.(\ref{sd1action}),(\ref{seis}),(\ref{diez})and (\ref{cotton}) respectively.
 The form of the actions appear in (\ref{sd1action}), (\ref{sd2action}), (\ref{sd3action2}) and (\ref{sd4action}).

There are several extensions of this work that could be
considered. It would be interesting to determine how these models
behave when interactions are included. Also, one could analyze
whether the dualities we established can be also found in the
gravity models that are connected via quadratic linearization with
the SD2, SD3 and SD4 models~\cite{VCS},
\cite{DJT},\cite{andringa}-\cite{dalmazinew}. Since there are
quadratic linearizations of topologically massive
supergravity~\cite{Deser:1982sw} and new topologically massive
supergravity~\cite{andringa}, it could be of interest to study the
possible equivalence between these models. Finally
 the extension of the quantum duality connections to the case of
higher spin theories~\cite{Aragone:1993jm}-\cite{Tyutin:1997yn}
which have self-dual formulations for massive excitations. We hope to discuss
 these issues in a future work.

 \vspace{1 cm}

\noindent\underline{Acknowledgments}
This work is partially supported by project PI-03- 00-5753-04 of
CDCH-UCV, Venezuela and PIP1787-CONICET,  PICT20204-ANPCYT grants and by CIC and UNLP,
Argentina. PJA would like to thank the hospitality and financial
support while his visit to the Department of Physics of the UNLP
where this work was started.

\end{document}